# Intra-Pulse Intensity Noise Shaping by Saturable Absorbers

MARVIN EDELMANN,[1,2,*] MIKHAIL PERGAMENT,[1] AND FRANZ X. KÄRTNER[1,2]

[1] Center for Free-Electron Laser Science CFEL, Deutsches Elektronen-Synchrotron DESY, Notkestr. 85, 22607 Hamburg, Germany
[2] Department of Physics, Universität Hamburg, Jungiusstr. 9, 20355 Hamburg, Germany
*Corresponding author: marvin.edelmann@desy.de

In this work, we identify and characterize intra-pulse intensity noise shaping by saturable absorbers applied in mode-locked lasers and ultra-low noise nonlinear fiber amplifiers. Reshaped intra-pulse intensity noise distributions are shown to be inevitably interconnected with self-amplitude modulation, the fundamental physical mechanism for initiation and stabilization of ultra-short pulses in the steady-state of a mode-locked laser. A theoretical model is used to describe the ultrafast saturation dynamics by an intra-pulse noise transfer function for widely-applied slow and fast saturable absorbers. For experimental verification of the theoretical results, spectrally-resolved relative intensity noise measurements are applied on chirped input pulses to enable the direct measurement of intra-pulse noise transfer functions using a versatile experimental platform. It is further demonstrated, how the characterized intra-pulse intensity noise distribution of ultrafast laser systems can be utilized for quantum-limited intensity noise suppression via tailored optical bandpass filtering.

## 1. Introduction

Saturable absorbers (SA) are key devices in mode-locked lasers, leading to short pulse formation and stabilization. In particular the ultra-low noise performance of mode-locked lasers in terms of pulse-to-pulse intensity-fluctuations, phase-noise, center frequency jitter and timing-jitter are of major importance for many cutting-edge applications including nonlinear microscopy [1], frequency-metrology [2] and photonic microwave extraction [3]. The significance of SA's with respect to laser mode locking has been amply studied in the past, in particular due to their critical role for self-amplitude modulation (SAM), the physical mechanism fundamentally enabling the pulse formation in mode-locked lasers and the stabilization of versatile mode-locked steady-states [4]. Numerous theoretical and experimental studies, including the pioneering work of Haus et al. and others [5-9], have identified SA's in their interaction with the circulating intra-cavity pulse via SAM as major contributor to the overall performance of mode-locked lasers, particular decisive for their noise behavior and parameter boundaries [10-13]. In this context, the concept of inverse saturable absorption has been demonstrated as a possible way to intrinsically stabilize the laser dynamics in mode-locked steady-states against q-switching via nonlinear power limiting resulting from an operation of SA's in an oversaturated regime [14-17]. Further elucidating the significant influence of SA working points for the noise performance of mode-locked lasers in accordance with these previous results, we recently demonstrated a nonlinear fiber amplifier system where the intentional operation of an artificial Kerr-type SA in such a strongly oversaturated regime enabled significant suppression of the relative intensity noise (RIN) down to the shot noise limit under proper conditions [18]. This RIN transfer mechanism of SA's has further shown to be the driving source for steady-states with strong intrinsic intensity noise suppression or enhancement in state-of-the-art fiber oscillators mode-locked with nonlinear amplifying loop mirrors (NALM) [19].

In this work, we present a comprehensive explanation for these observations by demonstrating that the nonlinear saturation dynamics and resulting SAM of slow and fast SA's are inevitably interconnected with an intrinsic ultrafast mechanism that strongly re-shapes intra-pulse intensity noise distributions (IPID's). To describe this mechanism, a theoretical model is introduced for the numerical simulation of saturation dynamics and the resulting intra-pulse noise transfer of SA's. To further demonstrate the IPID shaping of SA's experimentally, a versatile experimental platform is constructed that enables the direct measurement of intra-pulse noise transfer functions on the example of widely-applied slow and fast SA's, with a technique based on spectrally-resolved RIN measurements on chirped input pulses. With systematic IPID transfer measurements on a slow semiconductor saturable absorber mirror (SESAM) [20-22] and a fast Kerr-type nonlinear fiber interferometer (NLI) [23,24], the IPID shaping mechanism is experimentally verified and clearly associated with the nonlinear saturation dynamics in the respective SA, demonstrating good agreement with the numerical results and theoretical trends.

We further investigate the relation between the characterized IPID and the overall measurable RIN of the optical pulse train at the system output. By optically filtering local IPID minima in the spectral RIN distribution with tailored bandpass-filters at the output of an ultrafast laser system, we demonstrate shot-noise limited suppression of the RIN spectral density above 100 kHz offset-frequency. In agreement with this result, recent experimental studies by Kim et al. demonstrated and investigated the suppression of RIN and timing-jitter in mode-locked fiber lasers via narrow optical bandpass filtering [25,26].

## 2. Mechanism and Numerical Simulations

To describe and investigate the IPID shaping of SA's in an experimentally accessible parameter space, a model is developed and applied to compute the SA's intra-pulse RIN transfer functions. Intensity fluctuations are included via the definition of a fluctuating complex field $A(t)$ expressed with the equation

$$A(t) = \bar{A}(t) + \delta A(t) \quad (1)$$

composed of the average field component $\bar{A}(t)$ and a time-variant noise term $\delta A(t)$. With this expression, one can define

$$RIN(t) \equiv |\delta A(t)|^2/|\bar{A}(t)|^2 \quad (2)$$

as the time-dependent intra-pulse (relative) intensity noise distribution (IPID) of the time-domain pulsed input field $|A(t)|^2$. To compute the IPID shaping of a given optical system, we further define an intra-pulse RIN transfer function $H(t)$ according to the relation

$$H(t) \equiv \frac{RIN_{out}(t)}{RIN_{in}(t)} = \frac{|\delta A_{out}(t)|^2 |\bar{A}_{in}(t)|^2}{|\delta A_{in}(t)|^2 |\bar{A}_{out}(t)|^2} \quad (3)$$

with the fluctuating pulse intensity $|A_{in}(t)|^2$ at the system input and $|A_{out}(t)|^2$ at the output. For the case of SA's used in mode-locked lasers with known steady state intensity $|A_{in}(t)|^2$, the output field $|A_{out}(t)|^2$ can be obtained from its nonlinear power transfer function

$$H_P(t) \equiv \frac{|A_1(t)|^2}{|A_0(t)|^2} = 1 - q_s(t) \quad (4)$$

with the time-dependent SA saturation $q_s(t)$. To calculate the response of the SA-specific $q_s(t)$ to the fluctuation of $|A_{in}(t)|^2$, we use the well-known differential saturation equation

$$\frac{dq_s(t)}{dt} = -\frac{q_s(t) - q_0}{\tau_A} - q_s(t)\frac{|A_0(t)|^2}{E_{sat}} + d_{SA}\beta\frac{d|A_0(t)|^2}{dt} \quad (5)$$

where $q_0$ denotes the non-saturated loss, $\tau_A$ the absorber recovery time, $E_{sat}$ the saturation energy, $d_{SA}$ the thickness of the absorber layer and $\beta$ the two-photon absorption (TPA) coefficient [27,28].
The numerical procedure for computing $H(t)$ to quantify the IPID shaping for a given SA is given in the following. At first, the system response to the average field $\bar{A}_{in}(t)$ is simulated by solving Eq. (5) with the 4th order Runge-Kutta algorithm, yielding the average output field $\bar{A}_{out}(t)$. Next, the total output field $A_{out}(t)$ is determined including its fluctuation component by propagating $A_{in}(t)$ with added intra-pulse noise term $\delta A_{in}(t)$ (defined or measured) through the same system. The subsequent calculation of the output noise term $\delta A_{out}(t) = A_{out}(t) - \bar{A}_{out}(t)$ then allows to determine the output intra-pulse $RIN_{out}(t)$ and the SA intra-pulse noise transfer function $H(t)$ according to Eq. (3).

With respect to the experimental part of this work, a description of the IPID transfer function in the frequency-domain is essential. For this, Eq. (2) can be applied to the fluctuation input spectrum $\tilde{A}(\omega) = \bar{A}(\omega) + \delta A(\omega)$ (corresponding to the Fourier-transform of Eq. (1)), to obtain the spectral intra-pulse $\widetilde{RIN}_{in}(\omega)$ of the input field. Subsequently, Eq. (5) is applied in the time domain to include the ultrafast saturation dynamics of the SA by using the previously described numerical sequence, resulting again in the output field $A_{out}(t)$. Applying the Fourier-transformation to the time-domain output signals $\bar{A}_{out}(t)$ and $\delta A_{out}(t)$ then yields the frequency-domain average component $\bar{A}_{out}(\omega)$ and the noise component $\delta A_{out}(\omega)$. Subsequently, the frequency-resolved output $\widetilde{RIN}_{out}(\omega)$ can be calculated together with the spectral intra-pulse noise transfer function $\tilde{H}(\omega)$ of the SA, defined as

$$\tilde{H}(\omega) \equiv \frac{\widetilde{RIN}_{out}(\omega)}{\widetilde{RIN}_{in}(\omega)} = \frac{|\delta A_{out}(\omega)|^2 |\bar{A}_{in}(\omega)|^2}{|\delta A_{in}(\omega)|^2 |\bar{A}_{out}(\omega)|^2} \quad (6)$$

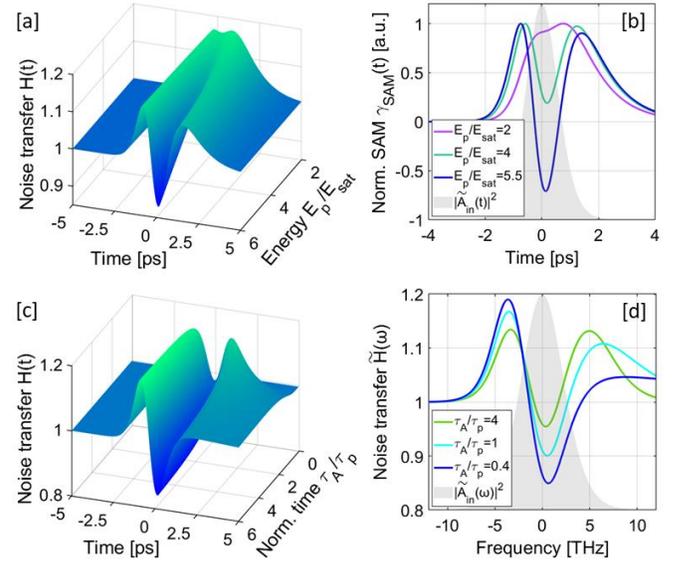

**Fig.1:** [a]: Simulated time-domain intra-pulse RIN transfer $H(t)$ as function of the norm. pulse energy $E_p/E_{sat}$. [b]: Corresponding normalized instantaneous SAM coefficient $\gamma_{SAM}(t)$ for selected values of $E_p/E_{sat}$, compared to the input pulse $|A_{in}(t)|^2$. [c]: Simulated $H(t)$ as function of the normalized SA recovery time $\tau_A/\tau_p$. [d]: Corresponding frequency domain $\tilde{H}(\omega)$ for selected values of $\tau_A/\tau_p$, compared to the input spectrum $|\tilde{A}_{in}(\omega)|^2$.

As a first step, we apply the numerical model to investigate key characteristics of the IPID shaping with SAs and to demonstrate its interconnection with the SAM in mode-locked lasers. For the following simulations, the SA parameters in Eq. (5) are set to $E_{sat} =$

$1\ nJ$, $\tau_A = 1\ ps$ and $q_0 = 0.55$, representing a typical parameter space e.g., for mode-locked fiber lasers [27]. The TPA parameters are matched to GaAs-based SESAM devices, with $d_{SA} = 2\ \mu m$ and $\beta = 2.5 * 10^{-10}\ m/W$. The Gaussian-shaped and positively chirped input pulse $|A_{in}(t)|^2$ has a full-width at half maximum (FWHM) spectral bandwidth of $5.3\ THz$, a group delay dispersion (GDD) of $-0.025\ ps^2$ and $\tau_p = 1.3\ ps$. It should be noted that the input pulse shape has a critical influence on the shape of the intra-pulse noise transfer function H(t) per definition given by Eq. (3). The initial noise term $\delta A_{in}(t)$ is set to ensure a constant RIN of 0.01% for each discrete time-segment of the total input pulse $|A_{in}(t)|^2$. Fig.1 [a] shows the calculated $H(t)$ in time-domain as a function of the normalized pulse energy $E_p/E_{sat}$. As shown, shape and magnitude of the simulated $H(t)$ changes significantly for variations of $E_p/E_{sat}$. While an overall RIN amplification can be observed for comparable low pulse energies in the range of $E_p/E_{sat} \leq 4.5$, larger values of $E_p/E_{sat}$ result in RIN amplification at the edges of $|A_{in}(t)|^2$ with simultaneously strong RIN suppression in its center. As demonstrated in Fig.1 [b]. The origin of this specifically varying $H(t)$ within $|A_{in}(t)|^2$ can be found in the SAM defined as the derivative of the SA reflectance with respect to the instantaneous pulse intensity, in other words by the varying slope of energy transfer proportional to the intensity within the duration of $|A_{in}(t)|^2$. Here, the SAM is quantified via the coefficient $\gamma_{SAM}(t) = dR(t)/dP(t)$, where $R(t) = 1 - q_s(t)$ denotes the SA reflectance and $P(t)$ the instantaneous pulse power. Sufficiently large values of $E_p/E_{sat}$ result in a changing sign of $\gamma_{SAM}(t)$ around the peak of $|A_{in}(t)|^2$ due to a growing influence of the TPA term in Eq. (5), interconnected with the well-known "roll-over" of the well-known intensity-dependent saturation curve of the SA [29].

To investigate the IPID shaping with respect to possible differences arising from slow and fast SA dynamics, Fig.1 [c] shows $H(t)$ as function of the normalized SA recovery time $\tau_A/\tau_p$. The simulation parameters are similar to the ones previously defined; however, the pulse energy is set to $E_p = 0.5\ nJ$ and the SA saturation energy to $E_{sat} = 3.5\ nJ$. For $\tau_A \ll \tau_p$ (characteristic for fast SA's) $H(t)$ has a symmetrical shape with RIN amplification at the front and back of $|A_{in}(t)|^2$, that decays towards the pulse center due to progressing saturation of the SA corresponding to a reduced SAM. For $\tau_A \gg \tau_p$ (slow SA), the SAM approaches zero in the back of $|A_{in}(t)|^2$ with a corresponding decay of $H(t)$, due to the progressive inability of $q_s(t)$ to recover within the time window of $\tau_p$ which severely reduces the saturation dynamics of the SA. In the pulse front, the progressively rapid saturation of the SA for higher intensities results in a comparable stronger SAM and RIN amplification. In the center of $|A_{in}(t)|^2$ an increasing RIN suppression can be observed, as consequence of the enduring saturation for larger $\tau_r$, resulting in a relative increase of the TPA contribution to the SAM according to Eq. (5).

An experimental relevant characteristic of the IPID shaping arises from the applied linear chirp of $|A_{in}(t)|^2$. As shown in Fig.1 [c] and [d], the resulting monotonously increasing instantaneous frequency establishes a linear relation between the shape of $H(t)$ and its Fourier-counterpart $\widetilde{H}(\omega)$ in the frequency-domain, respectively. To avoid the experimentally challenging task to directly measure $H(t)$ with femtosecond resolution, this relation allows to reconstruct shape and magnitude of the ultrafast IPID shaping from $\widetilde{H}(\omega)$ which is experimentally determinable via less demanding spectrally-resolved RIN measurements.

## 3. Experimental Setup and Methodology

To experimentally verify the IPID shaping of slow and fast SA's and to verify the numerical results, direct measurements of $\widetilde{H}(\omega)$ are performed on a SESAM and a Kerr-type NLI, respectively. Fig.2 [a] shows the experimental platform designed to enable these measurements. In the reference arm, a home-built Yb-doped fiber laser system generates a linearly polarized pulse train with 1030 nm center wavelength, 106 MHz repetition rate and a maximum pulse energy of ~1.1 nJ, corresponding to ~120 mW average power. A free-space isolator (ISO1) at the laser output prevents back-reflections and possible distortions of the laser operation. To determine $\tilde{H}(\omega)$ via frequency-resolved RIN measurements, the generated pulses are pre-chirped with a 1000 lines/mm transmission grating pair (LightSmyth T-1000-1040 Series). A polarization beam splitter (PBS1) combined with a half-wave plate (HWP1) is implemented as tunable output coupler for the reference signal $|A_1(t)|^2$ (subsequently, indices denote the system output port) to enable diagnostics and the measurement of its spectrally-resolved intra-pulse $\widetilde{RIN}_1(\omega)$. At PBS2, a second waveplate (HWP2) is used for an optional direction of $|A_1(t)|^2$ into one of two measurement arms of the experimental setup.

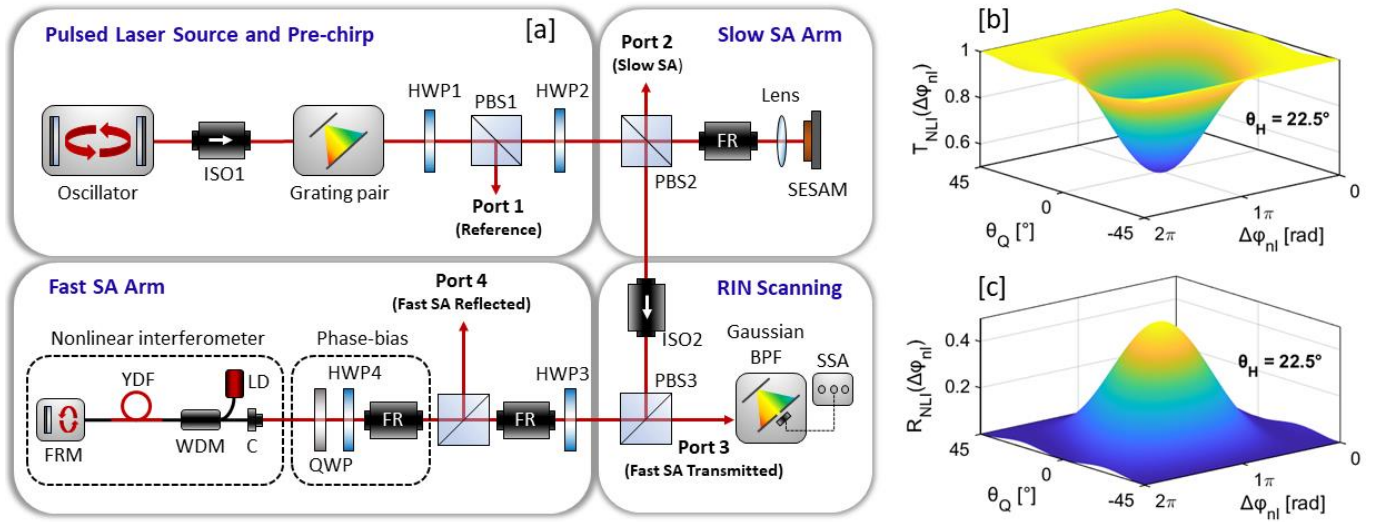

**Fig.2:** [a]: Experimental setup for the direct measurement of $\widetilde{H}(\omega)$ for a SESAM and NLI. ISO, isolator; HWP, half-wave plate; QWP, quarter-wave plate; PBS, polarization beam-splitter; SESAM, semiconductor saturable absorber mirror; FR, Faraday-rotator; C, collimator; LD, laser diode; WDM, wavelength division multiplexer; YDF, Ytterbium-doped fiber; FRM, Faraday-rotator mirror; BPF, bandpass filter; SSA, signal-source analyzer. [b]: NLI transmission function $T_{NLI}(\Delta\varphi_{nl})$ at Port 3 as function of the intensity-dependent differential nonlinear phase shift $\Delta\varphi_{nl}$ and the phase-bias QWP rotation angle $\theta_Q$ with $\theta_H$ fixed at 22.5°. [c]: Corresponding NLI reflectance $R_{NLI}(\Delta\varphi_{nl})$ at Port 4.

The slow SA arm contains a Faraday-rotator (FR, 45° single pass), a lens with 19 mm focal length and a SESAM (Batop SAM-1030-55-500fs). The double pass through the FR ensures a 90° polarization rotation of the reflected field $|A_2(t)|^2$ modulated by the SESAM dynamics, resulting in its emission at Port 2 for further measurements. With a saturation fluence of $F_{sat} = 50\ \mu J/cm^2$, $q_0 = 55\%$ and $\tau_A = 0.5\ ps$, the applied 19 mm focusing lens results in a maximum fluence on the SESAM of $F_{max} = 5 * F_{sat} = 250\ \mu J/cm^2$.

The fast SA arm consists of a NLI with free-space non-reciprocal phase bias (PB) attached to a polarization maintaining (PM) fiber segment. The fiber segment consists of ~0.5 m highly Yb-doped fiber (YDF, CorActive Yb-401 PM), optically pumped with a 1 W laser diode at 976 nm. The pump light is coupled into the YDF with a wavelength division multiplexer (WDM). Additional amplification with the YDF is required to ensure sufficient average power for measurements of $\widetilde{RIN}_{3,4}(\omega)$ at the NLI output ports by compensating the system losses of ~45%. The total fiber length of the NLI is ~1.8 m. The mechanism for SAM in this configuration is based on a differential nonlinear phase $\Delta\varphi_{nl}$, accumulated between the two orthogonal polarization modes in the PM-fiber via self-phase modulation [30,31]. Due to the double pass configuration with applied 90° Faraday-rotator mirror at one end, linear phase-shifts and the resulting drift between co-propagating pulses are compensated. The overlap in time with intensity-dependent $\Delta\varphi_{nl}$ results in a nonlinear polarization rotation, introducing a sinusoidal transmission function $T_{NLI}(\Delta\varphi_{nl})$ at PBS4 for the transmitted part (Port 3) and the reflectance function $R_{NLI}(\Delta\varphi_{nl}) = 1 - T_{NLI}(\Delta\varphi_{nl})$ for the reflected part (Port 4), which modulates the total nonlinearly rotated field that leaves the fiber segment of the NLI. The offsets and modulation depths of $T_{NLI}(\Delta\varphi_{nl})$ and $R_{NLI}(\Delta\varphi_{nl})$ are tunable via the settings of the non-reciprocal phase bias, consisting of a Faraday-rotator (FR), a half-wave plate (HWP4) and a quarter-wave plate (QWP). To calculate the transmission characteristics of the NLI, the Jones formalism can be applied consecutively for each polarizing component in the setup shown in Fig.2 [a], with an additional term that includes the nonlinear phase differences $\Delta\varphi_{nl}$ accumulated in the fiber amplifier. A detailed example of such analysis can be found in Ref. [32]. With this approach, one can describe and visualize the influence of the tunable half-wave plate HWP4 (rotation angle $\theta_H$) and the quarter-wave plate QWP (rotation angle $\theta_Q$) in the non-reciprocal phase bias on the intensity-dependent NLI transmission and reflection. Fig.2 [b] and [c] show the transmission $T_{NLI}$ as function of $\Delta\varphi_{nl}$ and the QWP rotation angle $\theta_Q$ for a fixed HWP4 rotation angle of $\theta_H = 22.5°$. The corresponding reflection $R_{NLI}$ with reciprocal shape and magnitude is shown in Fig.2 [c]. As indicated, the phase-bias rotation angles $\theta_Q$ and $\theta_H$ can be used to alter the state of the NLI transmission and reflection in terms of modulation depth and offset. In the context of mode-locked lasers, this can be used to optimize the SA settings for given boundary conditions of the cavity [30]. It should further be noted, that the observable reverse behavior of $T_{NLI}$ and $R_{NLI}$ results in an opposite sign between the SAM modulating the transmitted field $|\tilde{A}_3(\omega)|^2$ measurable at Port 3 and the SAM of the reflected field $|\tilde{A}_4(\omega)|^2$ at Port 4.

An essential part for the experimental verification of IPID shaping via SA's are consistent frequency-resolved RIN measurements. To ensure this, the RIN is determined via detection of the pulse-train with a fast and low-noise photodetector (ET-3000). Subsequently, the first harmonic at 106 MHz of the received RF-signal is filtered with a low-pass filter, amplified with a low-noise trans-impedance amplifier (MiniCircuits ZX60-33LN-S+) with stable power supply (Toellner TOE8721). To ensure a constant shot-noise level (SNL) at -142.3 dBc/Hz for all RIN measurement, the RF power is controlled via optical attenuation to maintain a constant value of -10 dBm. The double-sideband RIN spectrum of the RF signal is measured with the AM-noise function of a signal-source analyzer (SSA, Keysight E5052B) and integrated from 1 Hz to 10 MHz to obtain the integrated RIN value. To measure the frequency-resolved IPID given by $\widetilde{RIN}_N(\omega)$ at the respective output port N, a tunable gaussian bandpass-filter (GBPF) with 0.36 THz FWHM-bandwidth is constructed. As indicated in Fig.1 [a] at Port 3, the GBPF is realized with a transmission grating pair (1000 lines/mm), providing the diffraction-based spatial dispersion of the beam. Subsequently, a fiber-coupled collimator (Thorlabs, CFC-5X-B) with 4.6 mm focal length mounted on a linear translation stage is applied for tunable wavelength selection. With the GBPF, $\widetilde{RIN}_N(\omega)$ is measured by scanning through the respective output spectrum $|\tilde{A}_N(\omega)|^2$ with a step-size of 0.28 THz and measuring the RIN for each spectral segment. To ensure and RF power of > -10 dBm for each spectral segment over the whole 10 dB bandwidth of typical input spectra the following experiments, a minimum input power of ~80 mW is required at the entrance of the GBPF system.

## 4. Experimental Results and Discussion

### A. IPID Shaping of a slow SA (SESAM)

As a first experimental step, the IPID shaping of the SESAM is verified and characterized. To achieve this, its intra-pulse RIN transfer function $\tilde{H}(\omega)$ is measured as function of the normalized pulse duration $\tau_p/\tau_A$ of the reference pulse $|A_1(t)|^2$. The variation of $\tau_p/\tau_A$ with $\tau_A = 0.5\ ps$ is enabled via an adjustment of the pre-chirp through the grating distance. Fig.3 [a] shows the resulting AC traces of $|A_1(t)|^2$ for $\tau_p/\tau_A$ set to 10, 2.4 and 0.6 with a FWHM of 5 ps, 1.2 ps and 03 ps (assuming Gaussian pulse shape), respectively. With a constant pulse energy of $E_p = 0.95\ nJ$ (corresponding to ~5 * $E_{sat}$), the IPID of the reference pulse $|A_1(t)|^2$ is re-shaped by the saturation dynamics and nonlinear absorption of the SESAM, influenced by the varying $\tau_p/\tau_A$ and the corresponding step-wise increasing peak-power from ~0.17 kW for $\tau_p/\tau_A = 10$ to more than 2.8 kW for $\tau_p/\tau_A = 0.6$. Subsequently, the measured output spectra $|\tilde{A}_2(\omega)|^2$ shown in Fig.3 [b] are scanned with the GBPF to obtain the corresponding frequency-domain output IPID modulated by the SESAM, expressed via $\widetilde{RIN}_2(\omega)$. Fig.3 [c] shows the measured $\widetilde{RIN}_2(\omega)$ traces for different $\tau_p/\tau_A$, compared to the reference $\widetilde{RIN}_1(\omega)$ and $|\tilde{A}_1(\omega)|^2$. For $\widetilde{RIN}_1(\omega)$, two local minima can be observed close to the center frequency offset of $\Delta\omega = -11.6\ THz$ and $18.6\ THz$ with a modulation depth of the integrated RIN reaching from $0.097\%$ at $-23.8$ THz down to $0.074\%$ at $-11.6$ THz. Here, $\Delta\omega = \omega - \omega_c$ again denotes the offset of the 0.36 THz FWHM filtered spectral segments center-frequency to the center frequency $\omega_c$ of the respective spectrum $|\tilde{A}_N(\omega)|^2$.

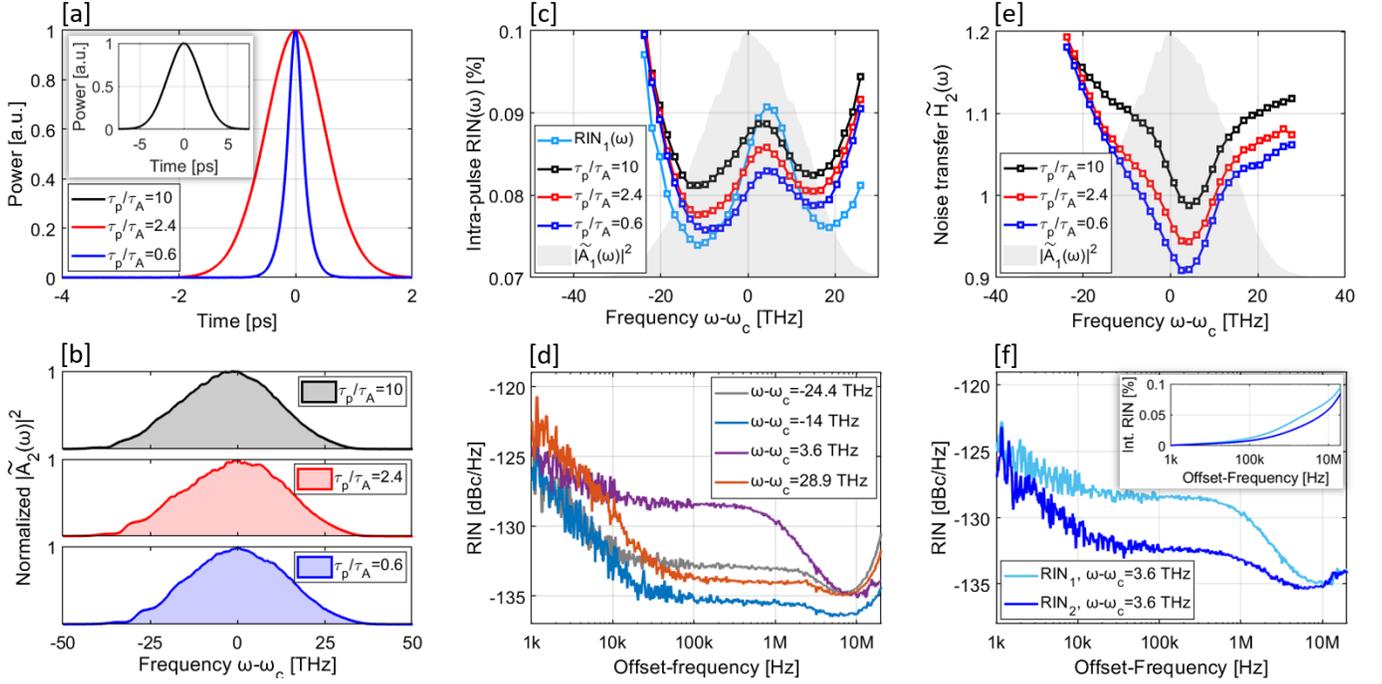

**Fig.3:** [a]: Measured autocorrelation traces of $|A_2(t)|^2$ at Port 2 for different values of $\tau_p/\tau_A$. [b]: Corresponding spectra $|\tilde{A}_2(\omega)|^2$. [c]: IPID traces $\widetilde{RIN}_2(\omega)$ (modulated by the SESAM) measured at Port 2 for different values of $\tau_p/\tau_A$, compared to the input IPID $\widetilde{RIN}_1(\omega)$ and $|\tilde{A}_1(\omega)|^2$. [d]: Measured RIN spectra of the input $\widetilde{RIN}_1(\omega)$ at characteristic frequencies $\omega - \omega_c$. [c]: Calculated $\tilde{H}(\omega)$ of the SESAM for different values of $\tau_p/\tau_A$, compared to $|\tilde{A}_1(\omega)|^2$. [d]: Offset-frequency resolved RIN spectral density measured at the filtered spectral segment centered at $\omega - \omega_c = 4.3\ THz$ of $|\tilde{A}_1(\omega)|^2$ (black) compared to the RIN spectrum of the corresponding spectral segment of $|\tilde{A}_2(\omega)|^2$ with $\tilde{H}(\omega) = 0.91$ (maximum noise suppression) at $\tau_p/\tau_A = 0.6$.

To illustrate the interconnection between the measured IPID and the corresponding offset-frequency resolved RIN spectral density of the filtered spectral segments as it is directly measured by the SSA, Fig.3 [d] shows the RIN spectra measured at characteristic values of $\Delta\omega$, corresponding to local minima or maxima of $\widetilde{RIN}_1(\omega)$. As shown, the varying shape and magnitude of $\widetilde{RIN}_1(\omega)$ is the result of significant and broadband intra-pulse variations of the RIN spectral density which is modulated by up to 8 dB between 10 kHz and 3 MHz offset-frequency. Fig.3 [e] shows the calculated $\tilde{H}(\omega)$ of the SESAM for varying $\tau_p/\tau_{SA}$, compared to $|\tilde{A}_1(\omega)|^2$. The shape of $\tilde{H}(\omega)$ indicates RIN amplification for the low and high frequency edges of $|\tilde{A}_2(\omega)|^2$, corresponding to the front and back of the chirped $|A_2(t)|^2$ in the time-domain, respectively. Towards the center of $|\tilde{A}_2(\omega)|^2$, the magnitude of $\tilde{H}(\omega)$ decays, indicating a less efficient RIN amplification shifting towards RIN suppression. In particular for a decreasing pulse duration $\tau_p/\tau_A$, stronger RIN suppression can be observed around the center of $|\tilde{A}_2(\omega)|^2$ with $\tilde{H}(\omega)$ reaching down to 0.91 for $\tau_p/\tau_A = 0.6$ at $\Delta\omega = 2.5\ THz$. In agreement with the simulations, this increasingly efficient RIN suppression close to the center of $|\tilde{A}_2(\omega)|^2$ (and thus close the peak of $|A_2(t)|^2$) for lower values of $\tau_p/\tau_A$ can be explained with the larger peak power and proportionally growing influence of TPA-induced oversaturation of the SESAM. The observed transition from RIN amplification at the front and back of $|\tilde{A}_2(\omega)|^2$ to RIN suppression around its center can be explained with the sign transition of the SAM that reaches increasingly negative values for lower $\tau_p/\tau_{SA}$ due to the growing TPA magnitude around the pulse peak power. The reduced magnitude of $\tilde{H}(\omega)$ at the back of $|A_2(t)|^2$ in time domain (high frequency edge of $|\tilde{A}_2(\omega)|^2$) can be explained with the influence of the SESAM recovery time $\tau_A$, which dampens the SAM magnitude and thus significantly hampers the dynamic saturation response and IPID shaping capability of the SESAM. For a normalized input pulse duration of $\tau_p/\tau_A = 0.6$, Fig.3 [f] shows the RIN spectral density measured on the spectral segment at $\Delta\omega = 2.5\ THz$ of $|\tilde{A}_2(\omega)|^2$ with maximum RIN suppression by $\sim 9\%$ ($\tilde{H}(\omega) = 0.91$) compared to the RIN spectrum of the corresponding spectral segment of the reference spectrum $|\tilde{A}_1(\omega)|^2$. As shown here, the saturation dynamics of the SESAM in an oversaturated regime with strong TPA magnitude around the peak of $|A_1(t)|^2$ introduce a broadband suppression of the RIN spectral density between 1 kHz and 10 MHz offset-frequency, reaching up to 7.5 dB in magnitude. In conjunction with the amplification of the integrated RIN around the edges of $|A_2(t)|^2$, significantly influenced by $\tau_A$, only a single interaction of the input field $|A_1(t)|^2$ with the saturation dynamics of the SESAM results in a strong re-shaping of the input IPID.

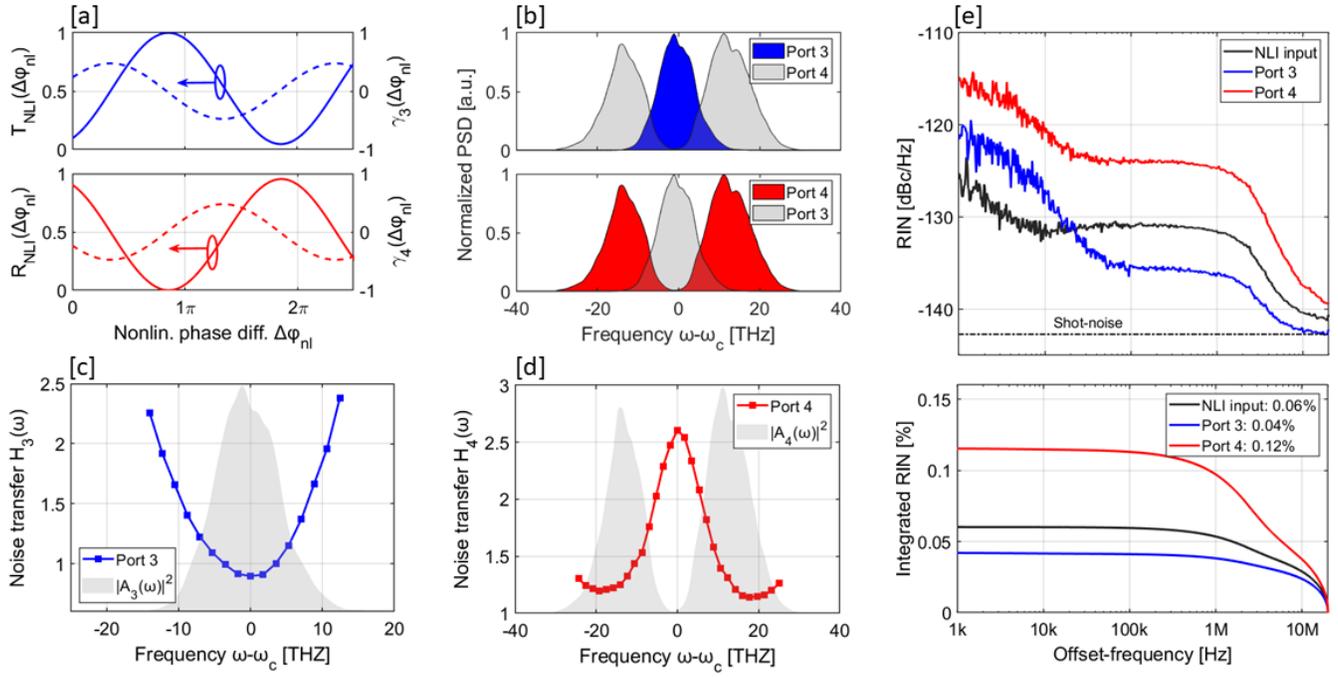

**Fig. 4:** [a]: NLI transmission $T_{NLI}(\Delta\varphi_{nl})$ (blue) at Port 3 and reflectance $R_{NLI}(\Delta\varphi_{nl})$ (red) at Port 4 with the corresponding SAM parameter $\gamma_T(\Delta\varphi_{nl})$ and $\gamma_R(\Delta\varphi_{nl})$ as function of the nonlinear phase difference $\Delta\varphi_{nl}$. [b]: NLI output spectra $|\tilde{A}_3(\omega)|^2$ and $|\tilde{A}_4(\omega)|^2$ measured at Port 3 and 4 of the NLI, respectively. [c]: Intra-pulse noise transfer function $\tilde{H}_3(\omega)$ measured on the NLI Port 3 output spectrum $|\tilde{A}_3(\omega)|^2$. [d]: Corresponding $\tilde{H}_4(\omega)$ measured on $|\tilde{A}_4(\omega)|^2$ at output port 4. [e] Measured RIN spectral density at the NLI input compared to the RIN at Port 3 and 4 with the corresponding integrated RIN from 1 kHz to 20 MHz.

**B. IPID Shaping of a fast SA (NLI)**

As next step, the IPID shaping of a fast SA is investigated at the example of the NLI implemented as part of the fast SA arm in the experimental platform in Fig1 [a]. To verify and investigate its intra-pulse noise transfer characteristics via measurements of the noise transfer function, $\tilde{H}(\omega)$, the chirped reference field $|A_1(t)|^2$ with spectrum $|\tilde{A}_1(\omega)|^2$ of similar shape to the one displayed in Fig.2 [b] and characterized $\widetilde{RIN}_1(\omega)$ is coupled into the NLI. The pre-chirp is set to ensure $\tau_p = 1\ ps$ and the pulse energy at the NLI input is fixed to $E_p = 0.6\ nJ$. To clearly verify the re-shaping of the input IPID via the fast intrinsic NLI saturation dynamics, $\widetilde{RIN}_3(\omega)$ and $\widetilde{RIN}_4(\omega)$ are measured influenced by the nonlinear transmission $T_{NLI} = 1 - q_s$ and the corresponding reflectance $R_{NLI} = q_s$ at Port 3 and 4, respectively. Calculated with the Jones-formalism, Fig.4 [a] shows $T_{NLI}(\Delta\varphi_{nl})$ and $R_{NLI}(\Delta\varphi_{nl})$ for the fixed experimental system working point with PB rotation angles of $\theta_Q = 112°$ and $\theta_H = -232°$ together with the respective SAM parameters $\gamma_T \propto dT_{NLI}(\Delta\varphi_{nl})/d\Delta\varphi_{nl}$ and $\gamma_R \propto dR_{NLI}(\Delta\varphi_{nl})/d\Delta\varphi_{nl}$ which shape the output field at Port 3 and 4, respectively. To ensure sufficient output power at both output ports for the frequency-resolved RIN scanning, the YDF in the NLI fiber segment is optically pumped with ~500 mW pump power from the LD, resulting in an average output power of 90 mW at Port 3 and 116 mW at Port 4. The nonlinear polarization rotation originated in the accumulation of the intensity-dependent $\Delta\varphi_{nl}$ ensures an inverse transmission behavior at Port 3 and 4 as demonstrated in Fig.4 [a], consequently correlated with opposing signs of $\gamma_T$ and $\gamma_R$ for each value of $\Delta\varphi_{nl}$. As shown in Fig.4 [b] the interconnection of $T_{NLI}(\Delta\varphi_{nl})$ and $R_{NLI}(\Delta\varphi_{nl})$ causes a distinct dependence with respect to the shape of the output spectra $|\tilde{A}_3(\omega)|^2$ and $|\tilde{A}_4(\omega)|^2$ separated at PBS3 from the total nonlinear modulated field leaving the NLI fiber segment. Fig.4 [c] and [d] show the measured intra-pulse noise transfer functions $\tilde{H}_3(\omega)$ and $\tilde{H}_4(\omega)$ with the corresponding output spectra at the NLI output Ports 3 and 4, respectively. As shown, $\tilde{H}_{3,4}(\omega)$ both show a symmetric trend with respect to the center of $|\tilde{A}_{3,4}(\omega)|^2$ at $\Delta\omega = 0\ THz$. Considering the numerical results for fast SA's, this symmetry can be explained with the instantaneous reaction of the non-resonant NLI saturation to $|A_1(t)|^2$, without any distortions as they occur for slow SA's e.g., due to the comparably long recovery time. A second key characteristic observable when comparing $\tilde{H}_3(\omega)$ with $\tilde{H}_4(\omega)$ is the clear inversion with respect to their magnitude and shape, similar to the one present in the output power spectral densities $|\tilde{A}_{3,4}(\omega)|^2$. At Port 3 with $|\tilde{A}_3(\omega)|^2$, significant RIN amplification can be observed in the fast and slow frequency range (corresponding to the front and back of $|A_3(t)|^2$) with $\tilde{H}_3(\omega)$ reaching more than 2.25 for $\Delta\omega < -12.5\ THz$ and $\Delta\omega > \sim 11\ THz$. Towards the center of $|\tilde{A}_3(\omega)|^2$, $\tilde{H}_3(\omega)$ monotonically decreases enabling RIN suppression by up to 10% at $\Delta\omega = 0\ THz$. In contrast, the filtered spectral segments at the slow and fast frequency edges of $|\tilde{A}_4(\omega)|^2$ consist of plateaus with local minima of $\tilde{H}_4(\omega)$ around $\Delta\omega = \pm 19\ THz$, reaching values of ~1.2. Towards the center of $|\tilde{A}_4(\omega)|^2$, monotonically increasing RIN amplification can be observed up to a maximum of $\widetilde{H}_4(\omega) = 2.6$ at $\Delta\omega = 0\ THz$.

In strong agreement with the theoretical evaluations, this inversely correlated shape and magnitude of $\widetilde{H}_3(\omega)$ and $\widetilde{H}_4(\omega)$ can be explained with the opposing sign of the energy transfer slope at Port 3 and 4 for each value of the intensity-dependent $\Delta\varphi_{nl}$, expressed via the respective SAM coefficients $\gamma_T(\Delta\varphi_{nl})$ and $\gamma_R(\Delta\varphi_{nl})$, respectively. To further elucidate the interconnection of the IPID shaping at the NLI output ports with the total RIN measured on the unfiltered output pulse trains, Fig.4 [e] shows the offset-frequency resolved average RIN spectral density measured Port 3 and 4 together with the integrated RIN (1 kHz to 20 MHz) compared to the RIN of the NLI input pulse train from the reference source. As shown, the IPID shaping of the NLI results in a significant variation of the measured total RIN with variations of the RIN spectral density of more than 10 dB between Port 3 and 4 in the offset-frequency range between 20 kHz and 10 MHz. Compared to the RIN of the NLI input measured with the reference signal $|A_1(t)|^2$, the IPID shaping of the NLI at Port 3 via $\gamma_T(\Delta\varphi_{nl})$ with RIN suppression around the peak of $|A_3(t)|^2$ enables a suppression of the total (unfiltered) RIN spectral density by up to ~5 dB in the high offset-frequency range > ~50 kHz, resulting in a corresponding suppression of the integrated RIN by 30% from 0.06% down to 0.04%, integrated from 1 kHz to 20 MHz. A possible explanation of the observable RIN spectral density amplification for offset-frequencies < ~20 kHz can be found in the additive pump noise transferred to the signal in the YDF. In agreement with our previous studies [18,19] where we consider only the transfer of the average RIN in NLI-based systems and fiber oscillators, the inverted $\gamma_R(\Delta\varphi_{nl})$ at Port 4 with significant RIN amplification around the center of $|A_3(t)|^2$ simultaneously results in a strong amplification of the total RIN spectral density by up to 6.7 dB with a corresponding amplification of the integrated RIN by 100% from 0.06% up to 0.12% (1 kHz to 20 MHz).

## C. RIN Suppression via Tailored Bandpass Filtering

To further elucidate the experimental capabilities arising from the interconnection of the IPID at the output of an ultrafast laser system with the measurable average RIN of the generated pulse train, the ability for maximum RIN suppression via tailored optical bandpass-filtering is investigated. For this experiment, the NLI phase-bias rotation angles are set to $\theta_Q = 245°$ and $\theta_Q = 42°$ with an input power of 40 mW and a YDF pump power of 400 mW, resulting in an output power of 133 mW and 66 mW at Port 3 and 4, respectively. In contrast to Ref. [18], the NLI phase-bias settings and pre-chirp are not adjusted for internal shot-noise limited (SNL) RIN suppression at Port 3. Instead, the system settings are chosen to ensure a measured $\widetilde{RIN}_3(\omega)$ at Port 3 with distinct local minima and maxima together with an average RIN at Port 3 above the SNL, as shown in Fig.5 [a] together with the corresponding output spectrum $|\tilde{A}_3(\omega)|^2$. Local minima of $\widetilde{RIN}_3(\omega)$ can be observed at $\Delta\omega = -17.5\ THz$ and $\Delta\omega = 0\ THz$ with the integrated RIN of the filtered segments taking values of ~0.05% and ~0.04%, respectively. In addition, two local maxima can be observed for the spectral segments centered at $\Delta\omega = -10.5\ THz$ and $\Delta\omega = -19.6\ THz$, corresponding to an integrated RIN of ~0.06% and ~0.08%, respectively. The resulting filtered spectra (BPF1-4) are shown in Fig.4 [b] compared to $\widetilde{RIN}_3(\omega)$ at NLI output Port 3.

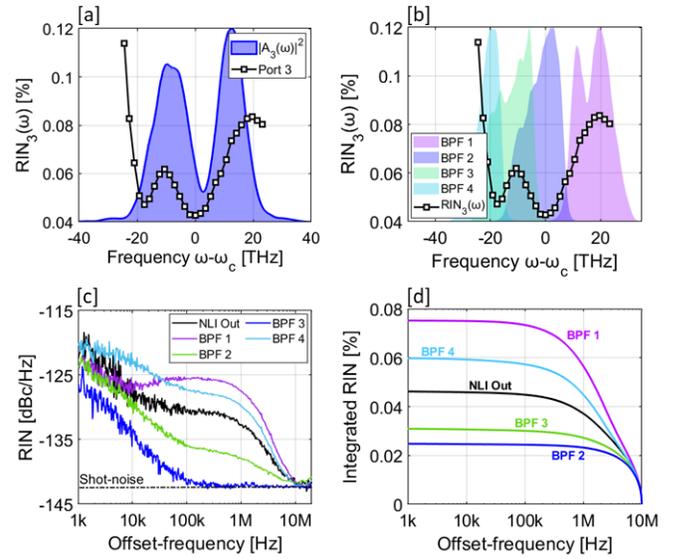

**Fig.5:** [a]: Measured $\widetilde{RIN}_4(\omega)$ at NLI output Port 4 with corresponding spectrum $|\tilde{A}_4(\omega)|^2$. [b]: $\widetilde{RIN}_4(\omega)$ with power spectral densities of filtered spectral segments (BPF1-4). [c]: Measured average RIN spectra for filtered spectral segments BPF1-4, compared to the average RIN spectrum of the NLI input. [d]: Corresponding integrated RIN in the range from 1 kHz to 10 MHz.

The corresponding RIN spectra measured for BPF1-4 are illustrated in Fig.4 [c] as function of the offset-frequency in the range from 1 kHz to 20 MHz and compared to the RIN spectrum of the NLI input signal $|A_1(t)|^2$. As shown, the application of BPF1-4 results in significant variations of the measured RIN spectral densities. By isolating the local IPID minimum at $\Delta\omega = 0\ THz$ with BPF4, shot-noise limited RIN suppression above 100 kHz by up to 12 dB is demonstrated with a remaining optical power of 50 mW. In contrast, spectrally isolating the local IPID maxima at $\Delta\omega = -10.5\ THz$ and $\Delta\omega = -19.6\ THz$ results in significant amplification of the average RIN by up to 5 dB above 100 kHz offset-frequency in the case of BPF1, with 90 mW filtered optical power. With this mechanism for RIN suppression and amplification via tailored bandpass filtering of local IPID extrema, recent results obtained by Kim et al. in Ref [25,26] can be explained, where significant improvement of the RIN and timing-jitter performance were experimentally demonstrated in state-of-the-art fiber oscillators via narrow optical bandpass-filtering.

## 5. Conclusion

In conclusion, we demonstrate ultrafast intra-pulse intensity noise shaping as a fundamental mechanism of saturable absorbers used in mode-locked lasers. To investigate this mechanism, a theoretical model is applied to compare the ultrafast saturation dynamics in slow and fast SA's via numerical simulations. It is found, that the interaction of intense laser pulses with SA's and the resulting self-amplitude modulation, the fundamental mechanism for pulse formation in mode-locked lasers, inevitably results in the shaping of significant intra-pulse intensity noise distributions as consequence of instantaneous variations of the energy transfer slope throughout the incident pulse. An experimental setup is constructed to directly measure the intra-pulse noise transfer functions of slow and fast

SA's with a technique of spectrally resolved RIN measurement with chirped input pulses. On the example of a semiconductor saturable absorber mirror and a Kerr-type nonlinear fiber interferometer, the IPID shaping is experimentally verified and investigated in good agreement with the theoretical results. It is further demonstrated, that the characterized IPID of an ultrafast laser system can be utilized to support quantum-limited RIN suppression or strong RIN amplification by filtering out local IPID extrema via tailored optical bandpass filtering.

**Funding.** Deutsche Forschungsgemeinschaft (KA 908/18-1), Deutsche Forschungsgemeinschaft - EXC 2056 - project ID 390715994

**Disclosures.** The authors declare no conflicts of interest.

**Data availability.** Data underlying the results presented in this paper are not publicly available at this time but may be obtained from the authors upon reasonable request.

## References

1. C. A. Casacio, L. S. Madsen, A. Terrasson, M. Waleed, K. Barnscheidt, B. Hage, M. A. Taylor, and W. P. Bowen, "Quantum-enhanced nonlinear microscopy," Nature **594**, 201–206 (2021).
2. J. Kim and Y. Song, "Ultralow-noise mode-locked fiber lasers and frequency combs: principles, status, and applications," Adv. Opt. Photon. **8**, 465-540 (2016).
3. X. Xie, R. Bouchand, D. Nicolodi, M. Giunta, W. Hänsel, M. Lezius, A. Joshi, S. Datta, C. Alexandre, and M. Lours, "Photonic microwave signals with zeptosecond-level absolute timing noise," Nat. Photonics **11**, 44–47 (2017).
4. H. A. Haus, "Mode-locking of lasers," IEEE J. Sel. Top. Quantum Electron. **6**, 1173–1185 (2000).
5. H. A. Haus, "Theory of Mode-Locking with a Fast Saturable Absorber," J. Appl. Phys. **46**, 3049–3058 (1975).
6. H. A. Haus, "Theory of Mode-Locking with a Slow Saturable Absorber," IEEE J. Quantum Electron. **11**, 736–746 (1975).
7. H. A. Haus and A. Mecozzi, "Noise of mode-locked Lasers," IEEE J. Quantum Electron. **29**, 983–996 (1993).
8. G. H. C. New, "Mode-locking of quasi-continuous lasers," Opt. Commun., vol. 6, p. 188, 1974.
9. E. P. Ippen, "Principle of passive mode locking," Appl. Phys. B **58**, 159–170 (1994).
10. R. Paschotta, "Timing jitter and phase noise of mode-locked fiber lasers," Opt. Express **18**, 5041-5054 (2010).
11. R. Paschotta, "Noise of mode-locked lasers (Part II): timing jitter and other fluctuations," Appl. Phys. B **79**, 163–173 (2004).
12. S. Namiki and H. A. Haus, "Noise of the stretched pulse fiber laser: part I—theory," IEEE J. Quantum Electron. **33**, 649–659 (1997).
13. R. Grange, M. Haiml, R. Paschotta, G. J. Spuhler, L. Krainer, M. Golling, O. Ostinelli, and U. Keller, "New regime of inverse saturable absorption for self-stabilizing passively mode-locked lasers," Appl. Phys. B **80**, 151-158 (2005).
14. T. R. Schibli, E. R. Thoen, F. X. Kärtner, and E. P. Ippen, "Suppression of Q-switched mode locking and break-up into multiple pulses by inverse saturable absorption," Appl. Phys. B **70**, S41-S49 (2000).
15. R. Grange, M. Haiml, R. Paschotta, G. J. Spuhler, L. Krainer, M. Golling, O. Ostinelli, and U. Keller, "New regime of inverse saturable absorption for self-stabilizing passively mode-locked lasers," Appl. Phys. B **80**, 151-158 (2005).
16. D. J. Harter, M. L. Shand, and Y. B. Band, "Power energy limiter using reverse saturable absorption," J. Appl. Phys. 56(3), 865–868 (1984).
17. D. J. Harter, Y. B. Band and E. Ippen, "Theory of mode-locked lasers containing a reverse saturable absorber," IEEE J. Quantum Electron. **21**, 1219-1228 (1985).
18. M. Edelmann, Y. Hua, K. Şafak, and F. X. Kärtner, "Nonlinear fiber system for shot-noise limited intensity noise suppression and amplification," Opt. Lett. **46**, 3344-3347 (2021).
19. M. Edelmann, Y. Hua, K. Şafak, and F. X. Kärtner, "Intrinsic amplitude-noise suppression in fiber lasers mode-locked with nonlinear amplifying loop mirrors," Opt. Lett. **46**, 1752-1755 (2021).
20. U. Keller, K. J. Weingarten, F. X. Kärtner, D. Kopf, B. Braun, I. D. Jung, R. Fluck, C. Hönninger, N. Matuschek, and J. A. D. Au, "Semiconductor saturable absorber mirrors (SESAM's) for femtosecond to nanosecond pulse generation in solid-state lasers," IEEE J Sel. Top. Quantum Electron. **2**, 435–453 (1996).
21. G. J. Spühler, K. J. Weingarten, R. Grange, L. Krainer, M. Haiml, V. Liverini, M. Golling, S. Schon, and U. Keller, "Semiconductor saturable absorber mirror structures with low saturation fluence," Appl. Phys. B **81**, 27-32 (2005).
22. I. D. Jung, F. X. Kärtner, N. Matuschek, D. H. Sutter, F. Morier-Genoud, Z. Shi, V. Scheuer, M. Tilsch, T. Tschudi, and U. Keller, "Semiconductor saturable absorber mirrors supporting sub-10-fs pulses," Appl. Phys. B **65**, 137–150 (1997).
23. M. E. Fermann, L.-M. Yang, M. L. Stock, and M. J. Andrejco, "Environmentally stable Kerr-type mode-locked erbium fiber laser producing 360-fs pulses," Opt. Lett. **19**, 43-45 (1994).
24. M. E. Fermann, F. Haberl, M. Hofer, and H. Hochreiter, "Nonlinear amplifying loop mirror," Opt. Lett. **15**, 752-754 (1990).
25. P. Qin, Y. Song, H. Kim, J. Shin, D. Kwon, M. Hu, C. Wang, and J. Kim, "Reduction of timing jitter and intensity noise in normal-dispersion passively mode-locked fiber lasers by narrow band-pass filtering," Opt. Express **22**, 28276-28283 (2014).
26. D. Kim, S. Zhang, D. Kwon, R. Liao, Y. Cui, Z. Zhang, Y. Song, and J. Kim, "Intensity noise suppression in mode-locked fiber lasers by double optical bandpass filtering," Opt. Lett. **42**, 4095-4098 (2017).
27. F. X. Kärtner, J. Aus der Au, and U. Keller, "Mode-locking with slow and fast saturable absorbers—what's the difference?" IEEE J. Sel. Top. Quantum Electron. **4**, 159–168 (1998).
28. M. Haiml, R. Grange, and U. Keller, "Optical characterization of semiconductor saturable absorbers," Appl. Phys. B **79**, 331–339 (2004).
29. E. R. Thoen, E. M. Koontz, M. Joschko, P. Langlois, T. R. Schibli, F. X. Kärtner, E. P. Ippen, and L. A. Kolodziejski, "Two-photon absorption in semiconductor saturable absorber mirrors," Appl. Phys. Lett. **74**, 3927-3929 (1999).
30. M. Edelmann, M. M. Sedigheh, Y. Hua, E. C. Vargas, M. Pergament, and F. X. Kärtner, "Large-mode-area soliton fiber oscillator mode-locked using NPE in an all-PM self-stabilized interferometer," Appl. Opt. **62**, 1672-1676 (2023)
31. X. Liu, R. Zhou, D. Pan, Q. Li, and H. Y. Fu, "115-MHz Linear NPE fiber laser using all polarization-maintaining fibers," IEEE Photonics Technol. Lett. 33, 81–84 (2021).
32. A. S. Mayer, W. Grosinger, J. Fellinger, G. Winkler, L. W. Perner, S. Droste, S. H. Salman, C. Li, C. M. Heyl, I. Hartl, and O. H. Heckl, "Flexible all-PM NALM Yb:fiber laser design for frequency comb applications: operation regimes and their noise properties," Opt. Express **28**, 18946-18968 (2020).